# Strain Induced Enhancement of Thermoelectric Properties of Monolayer WS$_2$ through Valley Degeneracy

Jayanta Bera, Satyajit Sahu*

Department of Physics, Indian Institute of Technology Jodhpur, Jodhpur 342037, India

**Abstract:**

    Two-dimensional transition metal dichalcogenides show great potential as promising thermoelectric materials due to their lower dimensionality, the unique density of states and quantum confinement of carriers. The effect of mechanical strain on the thermoelectric performances of monolayer WS$_2$ has been investigated using density functional theory associated with semiclassical Boltzmann transport theory. The variation of Seebeck coefficient and band gap with applied strain has followed the same type of trend. For n-type material the relaxation time scaled power factor($S^2\sigma/\tau$) increases by the application of compressive strain whereas for p-type material it increases with the application of tensile strain. A 77% increase in the power factor has been observed for the n-type material by the application of uniaxial compressive strain. A decrease in lattice thermal conductivity with the increase in temperature causes an almost 40% increase in ZT product under applied uniaxial compressive strain. From the study, it is observed that uniaxial compressive strain is more effective among all types of strain to enhance the thermoelectric performance of monolayer WS$_2$. Such strain induced enhancement of thermoelectric properties in monolayer WS$_2$ could open a new window for the fabrication of high-quality thermoelectric devices.

# Introduction:

Low dimensionality of materials opens a new window to enhance thermoelectric properties due to their unique density of states (DOS) and quantum confinement effect. Due to unique layered structure two dimensional (2D) materials have attracted much attention. Although the first synthesized 2D layered material graphene is well known for its very high mobility but opening a reasonable band gap in graphene is still a challenging problem[1]. Therefore 2D transition metal dichalcogenides (TMDCs) came into picture owing to its tunable bandgap, high electrical mobility, low thermal conductivity, chemical stability etc[2]. Fantastic combination of electrical and thermal transport in these 2D materials can lead to a significantly large thermoelectric figure of merit (ZT). TMDCs show a great potential in thermoelectric application to convert waste heat to electricity because of high electrical conductivity and low thermal conductivity. Theoretical calculation of thermoelectric properties of monolayer TMDCs suggests that ZT product of these materials are generally low[3] as compared to well-known thermoelectric materials such as $Bi_2Te_3$ and $Bi_2Se_3$[4]. Various efforts like chemical doping and functionalization[5], strain engineering[6], defect engineering[7], making heterostructures[8] have been taken into consideration to enhance the thermoelectric properties. Among these strain engineering is one of the most popular methods to tune the electronic and thermoelectric properties of these materials.

The effect of layer numbers on the thermoelectric power factor (PF) and ZT product has been predicted theoretically for $MoS_2$, $MoSe_2$, $WS_2$ and $WSe_2$ and the enhancement of ZT product in TMDCs is due to the increased degeneracy of the band edges[9,10] High ZT value close to 1 has been predicted in suspended monolayer and bilayer $MoS_2$ which is much higher than that of bulk suggests that low dimensionality may be a possible way to enhance the thermoelectric properties[11].By layer mixing or forming heterostructures of different TMDCs thermoelectric performance is enhanced due to increased DOS and reduction of band gap near the Fermi level [12,13]. Theoretical calculation predicted the value of lattice thermal conductivity of monolayer $MoS_2$ nanoribbon to be 29.2 W/m-K[14] which is higher than previous value of 1.35 W/m-K[15] but matched very well with experimental value of 34.5 W/m-K[16]. Experimental observation of very large power factor of 8.5 $mWm^{-1}K^{-2}$ at room temperature in exfoliated few layer $MoS_2$ because of the increased DOS and 2D confinement of electrons near the band edge suggests that 2D

TMDCs may turn out to be promising thermoelectric materials. Enhanced thermoelectric properties have been observed both experimentally and theoretically in monolayer $MoS_2$ with applied external electric field due to the change in valley degeneracy[17,18]. The effect of mechanical strain on the electronic and vibrational properties in different TMDCs has been investigated theoretically and suggests that TMDCs are very sensitive to mechanical strain[19–23] and monolayer $MoS_2$ is dynamically stable up to 15% biaxial tensile strain[24]. Enhancement of thermoelectric power factor by the application of compressive strain on monolayer $MoS_2$ has been predicted theoretically and maximum PF has been found for n-type doping with applied 3% uniaxial zigzag compressive strain[25]. A 2-4 % biaxial tensile strain results a reduction in $k_{ph}$ of monolayer suggests that strain engineering could be an effective way to increase ZT product[26].

Though in TMDCs, $MoS_2$ is the most widely studied material a thorough investigation of the role of strain on $WS_2$ is highly warranted. In this work to the best of our knowledge for the first time we have performed a systematic investigation of electronic, vibrational and thermoelectric properties of monolayer $WS_2$ and the effect of different types of strain on electronic and thermoelectric properties of monolayer $WS_2$. We have shown the variation of thermoelectric properties with chemical potential as well as carrier concentration and our study show that both the approaches lead to same results. The first principle calculation shows that the ZT product can be enhanced by 38.5% with applied uniaxial compressive strain and we have found the highest ZT product of 0.72 in monolayer $WS_2$. Such compressive strain induced enhancement of thermoelectric properties in monolayer $WS_2$ can have potential application in thermoelectric devices for the efficient conversion of wastage heat to electricity.

## Computational Details:

First principles calculation has been performed using density functional theory (DFT) with projector augmented wave (PAW)[27] potentials and the Perdew-Burke-Ernzerhof (PBE)[28] generalized gradient approximation (GGA)[29] as exchange correlation functional in Quantum Espresso (QE) package[30]. A sufficient vacuum of 17Å along C axis was created to avoid the interaction between layers in periodic boundary condition for monolayer of $WS_2$. A 24x24x1 dense mesh grid was used to optimize the geometry and energy cutoff for the electronic wavefunctions was set to 50 Ry throughout all the calculations. For the density of state (DOS) and thermoelectric parameter calculations a high dense mesh of K points 48x48x1 was used. The strain is calculated as $\varepsilon = \frac{|a-a_0|}{a_0} \times 100\%$ where $a_0$ and $a$ are the lattice constant of relaxed and strained structures respectively. For the thermoelectric and transport properties, a semi classical Boltzmann transport theory was used with constant scattering time approximation (CSTA) as implemented in BoltzTrap[31] code. In CSTA we assume that scattering time has very less dependency on energy and both the group velocity of carriers and DOS contribute to the transport function. The group velocity($v_g$) of carriers in a specific band can be described as

$$v_\alpha(i, \mathbf{k}) = \frac{1}{\hbar} \frac{\partial \epsilon(i, \mathbf{k})}{\partial \mathbf{k}_\alpha}, \qquad (1)$$

where $\mathbf{k}_\alpha$ is the αth component of wavevector $\mathbf{k}$ and $\epsilon(i, \mathbf{k})$ is the $i$th energy band and the conductivity tensor can be obtained in terms of group velocity as

$$\sigma_{\alpha\beta}(i, \mathbf{k}) = e^2 \tau(i, \mathbf{k}) v_\alpha(i, \mathbf{k}) v_\beta(i, \mathbf{k}), \qquad (2)$$

. The Seebeck coefficient, electrical conductivity and thermal conductivity due to electron can be calculated by using the values of group velocity $v_\alpha(i, \mathbf{k})$ as implemented in BoltzTrap[31] code by following equations

$$S_{\alpha\beta}(T, \mu) = \frac{1}{eT} \frac{\int v_\alpha(i, \mathbf{k}) v_\beta(i, \mathbf{k}) (\epsilon - \mu) \left[-\frac{\partial f_\mu(T, \epsilon)}{\partial \epsilon}\right] d\epsilon}{\int v_\alpha(i, \mathbf{k}) v_\beta(i, \mathbf{k}) \left[-\frac{\partial f_\mu(T, \epsilon)}{\partial \epsilon}\right] d\epsilon}, \qquad (3)$$

$$\frac{\sigma_{\alpha\beta}(T, \mu)}{\tau(i, \mathbf{k})} = \frac{1}{V} \int e^2 v_\alpha(i, \mathbf{k}) v_\beta(i, \mathbf{k}) \left[-\frac{\partial f_\mu(T, \epsilon)}{\partial \epsilon}\right] d\epsilon, \qquad (4)$$

$$\frac{k^{el}_{\alpha\beta}(T, \mu)}{\tau(i, \mathbf{k})} = \frac{1}{TV} \int v_\alpha(i, \mathbf{k}) v_\beta(i, \mathbf{k}) (\epsilon - \mu)^2 \left[-\frac{\partial f_\mu(T, \epsilon)}{\partial \epsilon}\right] d\epsilon, \qquad (5)$$

where $e, T, \tau, \mu, V$ are electronic charge, temperature, relaxation time, chemical potential, volume of an unit cell respectively and $f_\mu(T, \epsilon) = \frac{1}{e^{(\epsilon-\mu)/K_B T}+1}$ is the Fermi-Dirac distribution function.

For the calculation of lattice thermal conductivity due to phonon($k_{ph}$) we used Phono3py[32] code which calculates 2$^{nd}$ and 3$^{rd}$ order of force constant. The thermoelectric figure of merit (ZT) has been calculated using the formula

$$ZT = \frac{S^2 \sigma T}{k_{el} + k_{ph}}, \tag{6}$$

Where S, σ, T are Seebeck coefficient, electrical conductivity and temperature respectively and $k_{el}$ and $k_{ph}$ are thermal conductivity due to electron, and lattice thermal conductivity due to phonon respectively. The phonon limited carrier's mobility($\mu^{2D}$) and relaxation time(τ) has been calculated in another calculation(SI5.) with effective mass and deformation potential theory[33] by using the relation for 2D materials as follows

$$\mu^{2D} = \frac{2e\hbar^3 C_{2D}}{3K_B T |m^*|^2 E_{dp}^2} \quad \text{and} \quad \tau = \frac{m^* \mu}{e} \tag{7}$$

where m*, $C_{2D}$ and $E_{dp}$ are effective mass of carriers, stretching modulus and deformation potential respectively.

## Results and Discussions:

### Structural Parameters:

Monolayer WS$_2$ has a hexagonal honeycomb structure belonging to P63/mmc space group with S-W-S layer where W atoms are sandwiched between two layers of S atoms connected with covalent bonds known as 1H phase as shown in Fig. 1a. Bilayer WS$_2$ (2H) consists of two such S-W-S monolayers which are separated by Vander Waals interaction. The optimized lattice constant of a=b=3.19 Å for a unit cell of monolayer agrees with previous calculations [34] and experimental value[35]. To apply biaxial strain lattice parameter (a=b=3.19Å) is varied up to ±6% in the steps of 1% where + and – signs represent tensile and compressive strain respectively as shown in Fig. 1c. But in case of uniaxial strain we only vary lattice constant "a" up to ±6% while keeping b = 3.19Å fixed as shown in Fig. 1d.

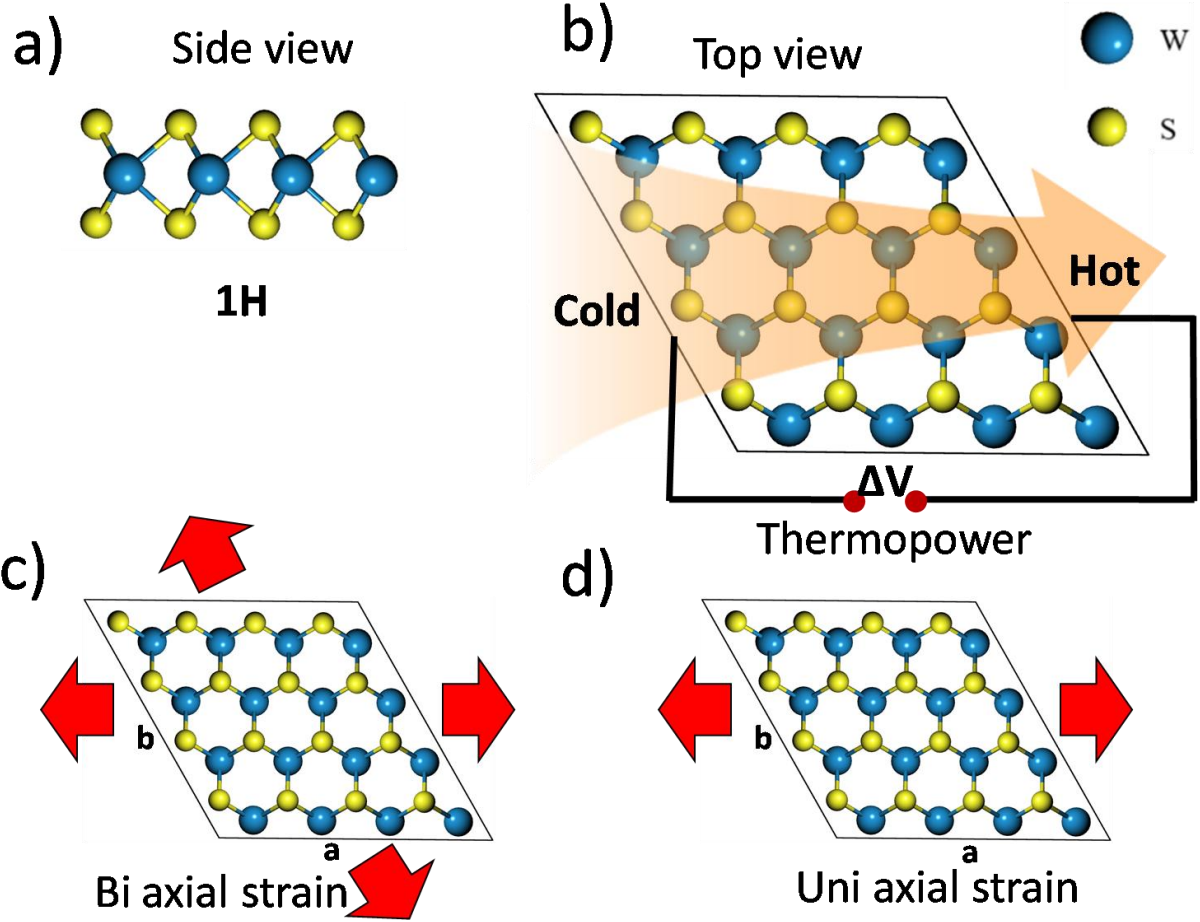

**Fig.1.** Crystal structure of monolayer WS$_2$ a) side view b) top view and thermoelectric effect c) application of bi axial strain d) application of uniaxial strain.

## Thermoelectric Properties:

Fig.2a shows the variation of carrier concentration of the WS$_2$ monolayer with chemical potential at three different temperatures namely 300K, 600K and 900K. Carrier concentration is zero in the band gap region and increases as we go to the band edge. The variation of seebeck coefficient (S) with chemical potential (µ) shows that very high S value has been observed near the bandgap region as shown on Fig.2b. S decreases as we go away from band gap region to the valance band and conduction band. This is because as we go away from bandgap region carrier concentration increases as shown in Fig.2a.and S is inversely proportional to carrier concentration. At 300 K the value of S is 2821 µV/K for p-type carriers and 2728µV/K for n-type carriers respectively in monolayer WS$_2$. The S values decreased to 1429 µV/K and1446 µV/K at 600K and 926 µV/K and 967 µV/K at 900K for p-type and n-type carriers respectively. At 300K electrical conductivity is almost zero in the region between the VBM and CBM but it increases at 600K and 900K as shown in Fig. 2c. This is due to the increase in the number of carriers at elevated temperatures. Because of high electrical conductivity the thermoelectric power factor ($S^2\sigma/\tau$) has also increased. Fig. 2d. shows how thermoelectric power factor $S^2\sigma/\tau$ (relaxation time scaled) varies with chemical potential µ(Ry). At 300 K the $S^2\sigma/\tau$ is observed to be $5.22 \times 10^{10}$ W/mK$^2$s for n-type carriers (µ>$E_f$) which increases with temperature and becomes $23.88 \times 10^{10}$ W/mK$^2$s at 900K. The highest $S^2\sigma/\tau$ at 300 K, 600 K, and 900 K are observed to be $8.57 \times 10^{10}$ W/mK$^2$s, $19.49 \times 10^{10}$ W/mK$^2$s, and $33.35 \times 10^{10}$ W/mK$^2$s for n-type region respectively. The corresponding $S^2\sigma/\tau$ at 300 K, 600 K, and 900 K for p-type carriers are $6.77 \times 10^{10}$ W/mK$^2$s, $12.90 \times 10^{10}$ W/mK$^2$s, and $18.4 \times 10^{10}$ W/mK$^2$s at VBM respectively which are lower than the values obtained for n-type carriers at CBM which indicates that n-type doping is more effective than p-type doping in semiconducting WS$_2$. At 300K in n-type region there are clearly three peaks in which first peak corresponds to the CBM. As temperature increases the two side peaks convert into two little hump and at 900K there is one peak corresponding to highest power factor. But no such thing is observed in p-type region. This type of behavior can be explained from DOS near the band edge which is shown in Fig. 2e. Near the valance band edge there is only one peak which corresponds to the energy state at VBM but in conduction band edge there are three humps near the conduction band edge and among them 1$^{st}$ hump corresponds to the CBM at 0.211 Ry and the second hump at 0.228 Ry corresponds to the highest $S^2\sigma/\tau$ value near the conduction band edge. As transport takes place near the band edge, we only check the peaks near the valance and conduction band edges. The variation of relaxation time scaled electrical conductivity i.e. $\sigma/\tau$ with temperature shows the semiconducting behavior of WS$_2$ and can be seen in Fig. 2f.

We have also shown the variation of Seebeck coefficient, $S^2\sigma/\tau$, $\sigma/\tau$ and $K_{el}$ as a function of carrier concentration (N) (SI Fig.S1.) and our results suggest that these values are similar to that calculated from chemical potential approach.

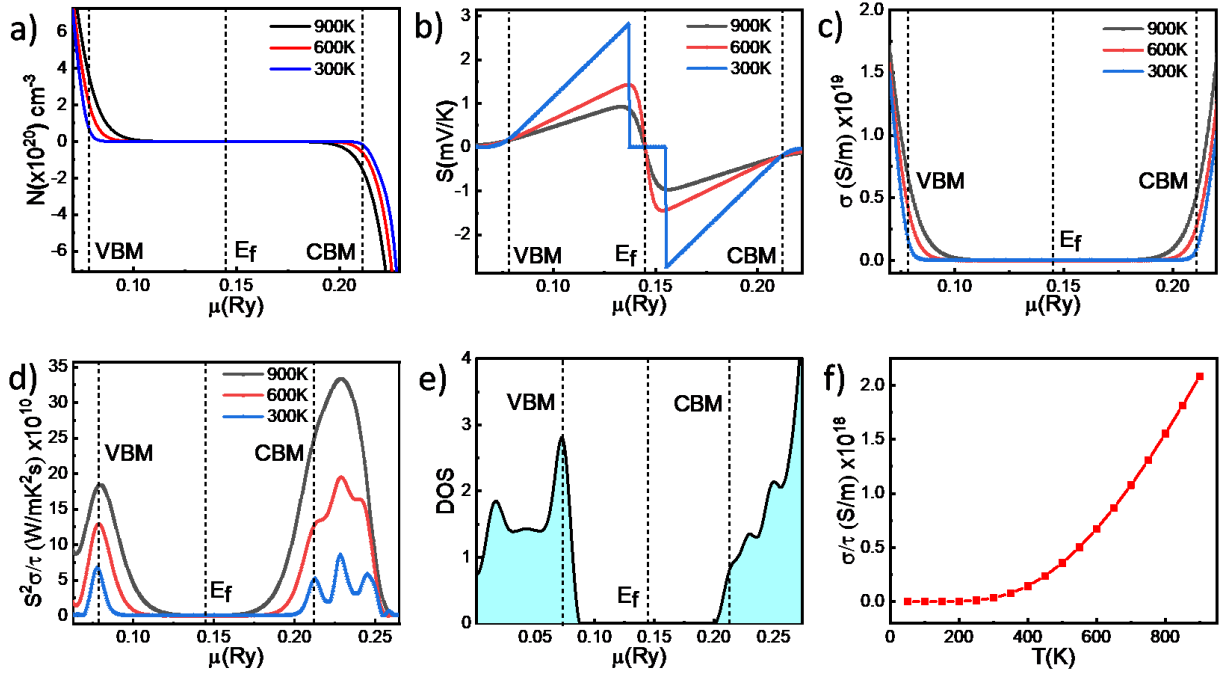

**Fig.2.** Variation of a) carrier concentration b) Seebeck coefficient(S) c) electrical conductivity (σ/τ) d) relaxation time scaled thermoelectric power factor ($S^2\sigma/\tau$) as a function of chemical potential μ(Rydberg unit). The dotted lines represent the valance band maximum(VBM), Fermi energy($E_f$) and conduction band minimum(CBM) respectively e) The density of states (DOS) as a function of chemical potential f) variation of electrical conductivity with temperature in monolayer $WS_2$.

### Effect of Bi-axial Strain:

The variation in Seebeck coefficient with applied biaxial strain (tensile and compressive) for n-type and p-type doping at 300K, 600K and 900K is shown in Fig.3a. In case of biaxial compressive strain (BCS) the value of S increases as we apply higher strain up to a point and after that point S starts decreasing. This behavior is observed for both p-type and n-type carriers, but S has higher value for n-type carriers than p-type carriers. The highest value obtained at 1% of BCS is 1067μV/K for n type and 955μV/K for p-type at 900K. But S decreases with application of biaxial tensile strain (BTS) for both types of carrier. The variation of band gap with applied strain follows the same trend as that of the Seebeck coefficient and is shown in Fig.3d. as described by Goldsmid-Sharp relation[36] given by the formula, $E_g = 2e|S_m|T_m$ where $S_m$ is the highest Seebeck coefficient, $T_m$ is the corresponding temperature and $E_g$ is the band gap. If we look at the electronic band structure of single layer WS$_2$ (SI Fig.2.b) it is clearly seen that at 1% BCS band gap is highest and remains direct and after 2% BCS it starts to decrease and becomes indirect. This is the reason for higher values of Seebeck Coefficient at 2% BCS. But for BTS band gap decreases and becomes indirect at 1% BTS and as BTS increases band gap decreases rapidly by going to the point of semiconducting to metal transition at 10% of BTS (SI Fig.S2.g). That is why Seebeck coefficient decreases by the application of BTS.

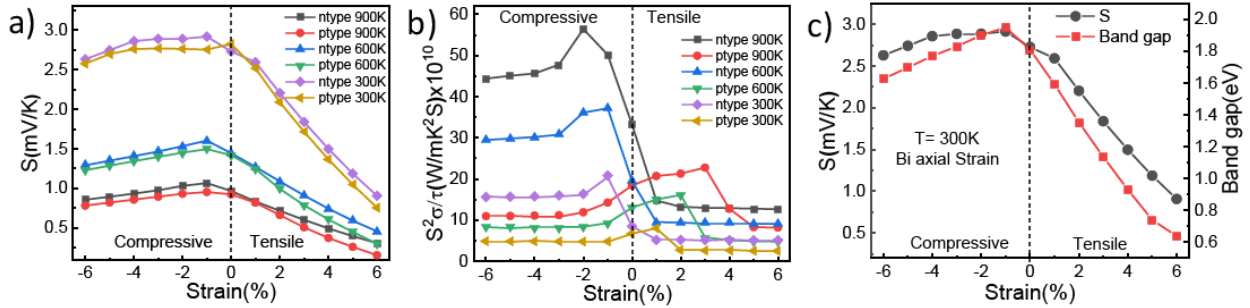

**Fig.3.** Variation in a) Seebeck coefficient and b) S$^2$σ/τ with applied both biaxial compressive and tensile strain for n-type and p-type doping at 300K, 600K and 900K. c) variation in Seebeck coefficient and band gap with the application of biaxial strain and those show similar type of trend.

The power factor S$^2$σ/τ (PF) increases significantly by applying BCS for n-type carriers and highest PF is found at 2% compressive strain with a value of 56.35 x 10$^{10}$ W/mK$^2$s which is more than 50% higher than that of without strain at 900K as shown in Fig.3b. However, by applying BCS the PF decreases up to 2% of BCS and then remains constant afterwards for p-type carriers.

An opposite trend was observed with application of BTS. In this case PF decreases rapidly with increasing tensile strain up to 2% and then becomes constant with a very lower value than that without strain for n-type carriers which suggest that application of BTS is not favorable for n-type doping. But thermoelectric PF increases for p-type carriers with the application of BTS. The highest PF is found to be 23.8 x 10$^{10}$ W/mK$^2$s which is 29% higher than that of without strain for p-type carriers at 3% of BTS and at 900K. So, at 900K PF is highest by applying 2% of BCS for n-type carriers but for p-type carriers PF has a maximum value by applying 3% of BTS though the maximum PF for n-type is much higher than that of the p-type.

From this analysis it is clear that BCS is most favorable for n-type doping whereas BTS is suitable for p-type doping.

## Effect of Uniaxial Strain:

The effect of uniaxial strain is slightly different from biaxial strain. Though the variation in Seebeck coefficient as shown in Fig.4a. does not differ very much than that of biaxial strain but the power factor variation is different. The Seebeck coefficient increases with uniaxial compressive strain (UCS) and the highest value is obtained at 2% UCS with a value of 1067μV/K for n-type and 955μV/K for p-type carriers at 900K which is exactly same that of 1% BCS. But S decreases with the application of uniaxial tensile strain (UTS) for both n-type and p-type carriers. Like biaxial strain here also S is higher for n-type than that of p-type carriers and the variations of S and band gap follow Goldsmid-Sharp relation. However, variation in $S^2\sigma/\tau$ (PF) shown in Fig.4b. is different from that of biaxial strain. The PF increases with applied UCS and attains the highest value of 59 x $10^{10}$ W/mK$^2$s at 4% of UCS, which is almost 77% higher than the value obtained without strain for n-type carriers. After 4% PF decreases but still remains at higher values than that of 0% strain. But with the application of UTS $S^2\sigma/\tau$ rapidly decreases for n-type carriers up to 4% of BTS and becomes constant with a much lower value than that without strain. It suggests that for n-type doping compressive strain is more effective than tensile strain.

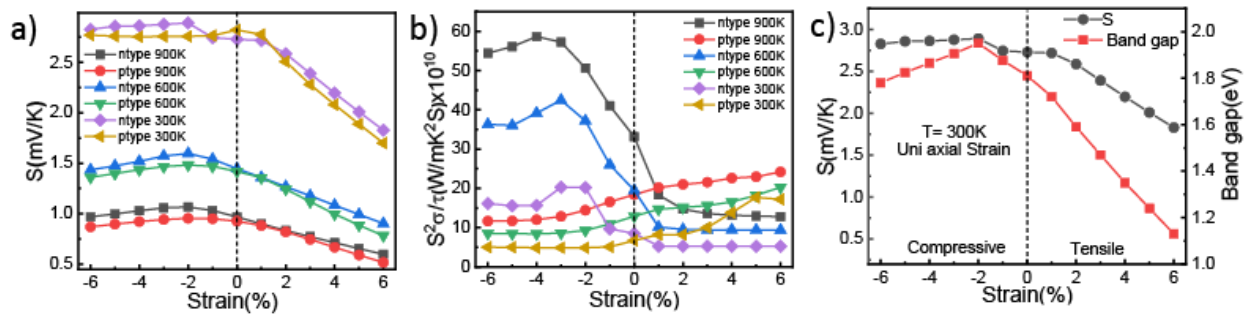

**Fig.4.** Variation in a) S and b) $S^2\sigma/\tau$ under the application of uniaxial compressive and tensile strain for n-type doping and p-type doping at 300K, 600K and 900K. c) Variation in S and band gap with applied uniaxial compressive and tensile strain.

For p-type carriers, $S^2\sigma/\tau$ decreases with application of UCS up to 4% and then becomes constant at a lower value than that without strain which is similar to that of BCS. But $S^2\sigma/\tau$ increases linearly with the application of UTS and attains a value of 24x $10^{10}$ W/mK$^2$s at 6% of UTS which is almost 30% higher than that without strain. So, the application of UCS affects more for n-type doping and UTS for p-type doping although the highest power factor at 4% of UCS for n-type is very much higher than that of p-type carriers.

**Electronic band structure and valley degeneracy:**

The electronic band structure of unstrained monolayer WS$_2$ is shown in Fig. 5a. Our calculation shows that the highest valance band has two hole valleys at Γ point (V1 and V3) and one at K point (V2) which are almost degenerate and the energy difference between the hole valleys, $\Delta V_{12}$ and $\Delta V_{23}$ are almost zero. In the conduction band three electron valleys at K point (C2), Λ point (C1) and the point in between Γ and M (C3) and they don't show valley degeneracy as the energy difference of $\Delta C_{12} = 0.2045$ eV and $\Delta C_{23} = 0.46$ eV are seen in Fig.5a. The conduction band minima (C2) and valance band maxima (V2) both lie on same K points and a direct band gap of 1.81 eV has been observed in unstrained monolayers $WS_2$. When we apply 1% biaxial compressive strain electron valleys at K point (C2) and Λ point (C1) become degenerate as difference between these two valleys vanishes ($\Delta C_{12} = 0$ eV) and $\Delta C_{23}$ reduces hence the conduction band shows valley degeneracy as shown in Fig. 5b. Now both direct (K-K) and indirect (K-Λ) transition can take place and both the electron valleys C1 and C2 contribute to the transport. This is why thermoelectric properties enhanced initially for n-type carries under the application of compressive strain. At the same time the hole valley at K point (V2) remains same but hole valleys at Γ point (V1 and V2) go downward and hole valley degeneracy in valance band is broken hence with the application of compressive strain thermoelectric performance of monolayer $WS_2$ for p-type carries decreases. As we increase biaxial compressive strain the valley degeneracy in conduction band as well as valance band are broken after 2% compressive strain and the thermoelectric performances decrease (SI Fig. S4.a). In the case of uniaxial compressive strain, the electron valley degeneracy occurs at 2% strain with a direct or indirect band gap of 1.96 eV and at this point hole valley degeneracy is broken as shown in Fig.5c. Electron valley degeneracy disappeared slowly at higher strain and after 4% uniaxial strain degeneracy breaks (SI Fig. S4.b).

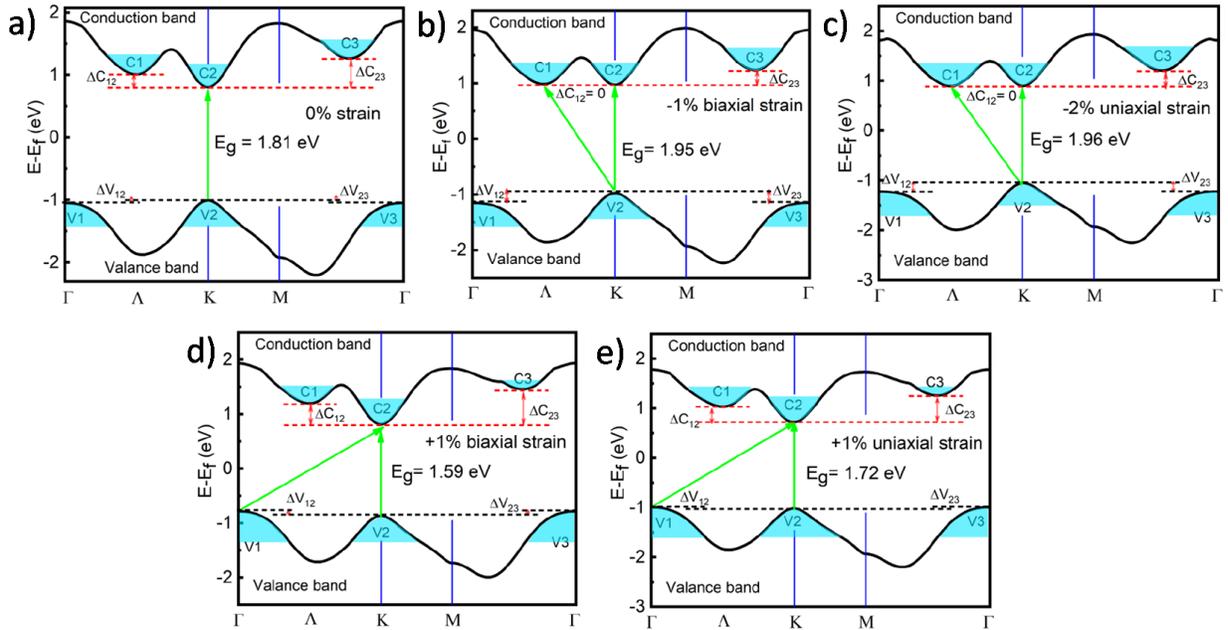

**Fig.5.** Band structure and valleys of monolayer WS$_2$ under the application of a) 0% strain b) -1% biaxial strain c) -2% uniaxial strain d) +1% biaxial strain e) +1% uniaxial strain. Here only the highest valance band and lowest conduction band is shown. The red dotted lines and black dotted lines represent energy levels of electron valleys in conduction band and hole valleys in valance band respectively. The green arrow represents direct/indirect band gap transition between hole and electron valleys.

As we apply 1% tensile strain (both biaxial and uniaxial) the hole valleys V1, V2, and V3 in valance band are nearly degenerate but the difference between conduction band valleys C1, C2 and C3 increases as shown in Fig.5d and Fig.5e. As CBM lies at K point an indirect (Γ-K) or direct transition (K-K) with a band gap of 1.59 eV and 1.72 eV for biaxial and uniaxial tensile strain respectively, can take place and both hole valleys at Γ point (V1 and V3) and K point (V2) can take part in transport. This is the reason thermoelectric power factor for the p-type doping initially increases with the application of tensile strain whereas thermoelectric properties for n-type doping decreases by tensile strain. Further enhancement of tensile strain results breaking of hole valley degeneracy very slowly (SI Fig.S4.c and S4.d) and the thermoelectric performances for p-type carriers also vary very slowly.

**ZT product:**

The variation in ZT product of unstrained monolayer $WS_2$ at 300K, 600K and 900K is shown in Fig.6a. We found that at 300K lattice thermal conductivity $(k_{ph})$ of unstrained monolayer $WS_2$ is 72 W/m-K. The variation of $k_{ph}$ with temperature is shown in Fig.6b. The highest ZT product of unstrained monolayer $WS_2$ has been found to be 0.52 for n-type carriers and 0.49 for p-type carriers at 900K. With applied BCS ZT product increases and highest ZT product with a value 0f 0.70 at 900K has been observed at 2% of BCS which is 35% higher than ZT value of unstrained $WS_2$ for n-type carriers. After 2% of BCS the variation is almost constant having a value of ZT = 0.62 at 6 % of BCS which is still19% higher than ZT of unstrained $WS_2$.

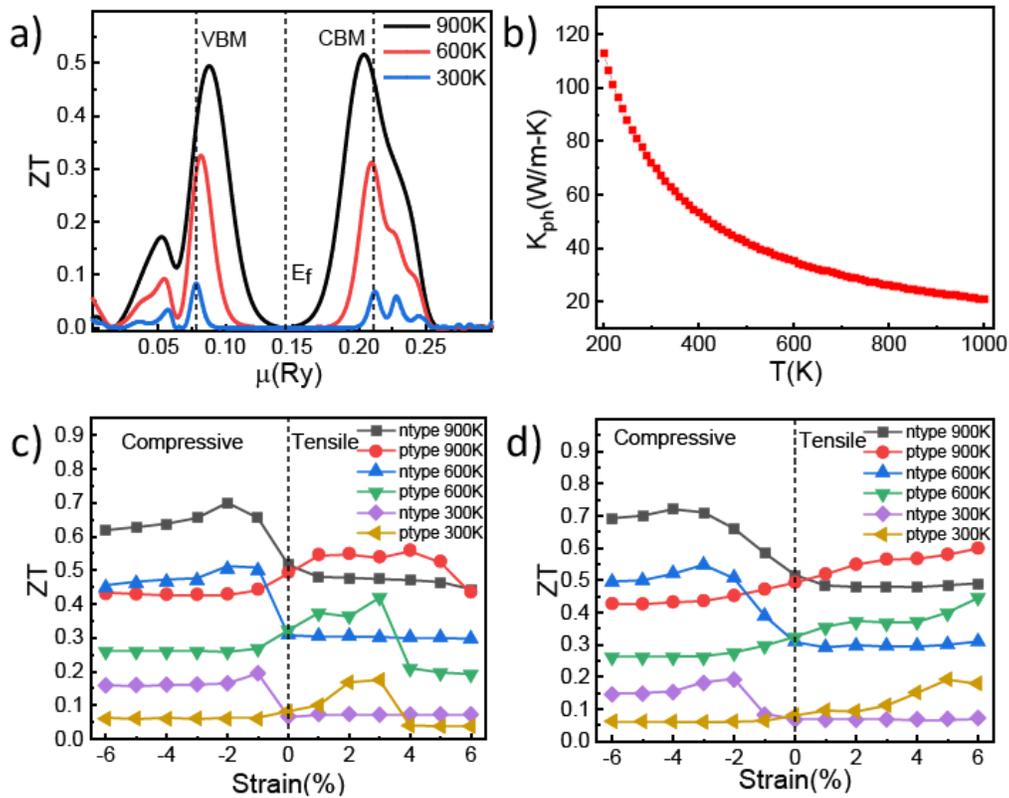

**Fig.6.** a) Variation of ZT product with chemical potential (μ) at 300K, 600K and 900K where VBM, CBM and Fermi level are shown by dotted lines and b) Variation of lattice thermal conductivity due to phonon $(k_{ph})$ with temperature in monolayer $WS_2$. Variation in ZT product with applied c) biaxial strain (compressive and tensile) d) uniaxial strain for n-type and p-type doping at 300K, 600K and 900K.

We have shown that monolayer WS$_2$ is thermodynamically stable below 4% of BCS(SI Fig. S5.d). But for the p-type carriers ZT products first decrease then becomes almost constant with the application of BCS. The variation of ZT values at different amount of bi axial strain and uniaxial strain has been shown in Fig. 6c. and 6d. at different temperatures. With the application of BTS ZT value for p-type carriers slightly increases up to 4% of BTS with a value of ZT= 0.56 and then decreases but for n-type carriers it does not vary much and slightly decreases at 900K.

The highest ZT products of ZT = 0.72 among all types of strain has been found for -4% uniaxial compressive strain (UCS) at 900K which is 38.5% higher than that of unstrained monolayer WS$_2$. The effect of UCS is almost similar to that of BCS but highest ZT has been found at 4%, 3%, and 2% of UCS at 900K, 600K and 300K respectively. Phonon dispersion curve at 4% of UCS (SI Fig. S5.e) shows monolayer WS$_2$ is stable at 4% of UCS. The effect of UTS is slightly different from that of BTS. For n-type carriers ZT value slightly decreases and then remains constant with the application of UTS but for p-type carriers ZT value changes very slowly up to 6% of UTS.

| Strain(%) | | S(μV/K) | | | S$^2$σ/τ (Wm$^{-1}$K$^{-2}$s$^{-1}$) | | | ZT | | |
|---|---|---|---|---|---|---|---|---|---|---|
| | | 300K | 600K | 900K | 300K | 600K | 900K | 300K | 600K | 900K |
| 0% | n | 2728 | 1446 | 968 | 8.50 | 19.50 | 33.15 | 0.067 | 0.308 | 0.516 |
| | p | 2821 | 1419 | 926 | 6.70 | 12.90 | 18.40 | 0.082 | 0.322 | 0.494 |
| -1% biaxial | n | 2914 | 1603 | 1067 | 20.82 | 37.15 | 50.10 | 0.195 | 0.500 | 0.655 |
| | p | 2751 | 1496 | 956 | 4.80 | 9.170 | 14.24 | 0.060 | 0.268 | 0.440 |
| -2% biaxial | n | 2887 | 1533 | 1032 | 16.35 | 36.10 | 56.35 | 0.165 | 0.504 | 0.700 |
| | p | 2760 | 1451 | 933 | 4.77 | 8.26 | 11.89 | 0.06 | 0.258 | 0.430 |
| +2% biaxial | n | 2204 | 1090 | 718 | 5.18 | 9.38 | 13.14 | 0.074 | 0.304 | 0.478 |
| | p | 2088 | 996 | 665 | 2.70 | 16.00 | 21.30 | 0.168 | 0.363 | 0.550 |
| -2% uniaxial | n | 2890 | 1597 | 1066 | 20.17 | 37.19 | 50.63 | 0.192 | 0.506 | 0.660 |
| | p | 2757 | 1483 | 954 | 4.86 | 9.28 | 14.42 | 0.0624 | 0.274 | 0.453 |
| -4% uniaxial | n | 2861 | 1519 | 1029 | 15.65 | 39.16 | 59.00 | 0.153 | 0.520 | 0.720 |
| | p | 2753 | 1436 | 922 | 4.86 | 8.41 | 12.00 | 0.062 | 0.262 | 0.433 |
| +4% uniaxial | n | 2195 | 1084 | 714 | 5.262 | 9.43 | 13.13 | 0.069 | 0.296 | 0.478 |
| | p | 2081 | 997 | 668 | 13.83 | 16.56 | 22.60 | 0.152 | 0.370 | 0.568 |

**Table 1.** Variation of S, S$^2$σ/τ and ZT with different type of strain at 300K, 600K and 900K for n-type and p-type carriers.

So, it is clear that ZT product increased to 38.5% at 4% UCS and 35% at 2% BCS than that of ZT value of unstrained monolayer $WS_2$. Also, from phonon dispersion curves at different strain it has been observed that monolayer $WS_2$ is thermodynamically stable at 4% of UCS and 2% of BCS. The value of Seebeck coefficient, power factor, and ZT product at different temperatures for different doped materials under both uniaxial and biaxial strain is shown in table-1.

## Conclusions:

In conclusion, we have performed strain dependent studies of electronic and thermoelectric properties of monolayer $WS_2$ by using DFT and Boltzmann transport theory. We have calculated thermoelectric properties as a function of chemical potential and carrier concentration and found that both the approaches give same output. The enhancement of the thermoelectric performances has been found with the application of compressive strain for n-type doping and that of tensile strain for p-type doping. Among all types of strains uniaxial compressive strain has been found to be most effective and highest thermoelectric PF and ZT product have been found to be 59 x $10^{10}$ W/mK$^2$s and 0.72 respectively at 900K under 4% of uniaxial compressive strain. The enhancement of thermoelectric properties is due to the change in degeneracy of bands near the Fermi level. These results clearly indicate that $WS_2$ could be a very promising material for thermoelectric applications under applied compressive strain.

## Acknowledgement:

The authors are thankful to Ministry of Human Resource and Development (MHRD) India for supporting this work. We are also thankful to Indian Institute of Technology Jodhpur (IITJ) for all the support to carry out the experiment.